\documentclass[aps,prl,twocolumn,amsmath,amssymb]{revtex4}%
\usepackage{amsfonts}
\usepackage{amsmath}
\usepackage{amssymb}
\usepackage{graphicx}%
\begin{document}
\title{Device modeling of long-channel nanotube electro-optical emitter}
\author{J. Tersoff$^{*}$, Marcus Freitag, James C. Tsang, and Phaedon Avouris}
\affiliation{IBM Research Division, T. J. Watson Research Center, Yorktown Heights, New
York 10598}
\date{\today}

\begin{abstract}
We present a simple analytic model of long single-wall
nanotube electro-optical emitters,
along with experimental measurements using improved devices
with reduced hysteresis.
The model describes well the voltage-controlled motion of the emission spot,
and provides a clear picture of the physical mechanism of device operation.
It also
indicates that the electric field is strongly enhanced at the emission spot,
and that device performance can be greatly improved by the use
of thinner gate oxides.

\end{abstract}
\maketitle

Carbon nanotube field-effect transistors (CNFETs) exhibit unique
electro-optical properties.
In particular, they can serve as electrically
pumped nanoscale light emitters \cite{Mis03}. Not only is the light source
localized to the width of tube ($\sim$1-2nm);  it is also localized along the
length of the tube, and can be \textit{electronically positioned}
\cite{Fre04b}. This suggests novel possibilities for electro-optical devices.
In addition, such devices provide a unique probe of the transport in
nanotubes, because the movable emission spot provides a direct measure of the
charge rearrangement in response to voltage changes.

Here we examine the device properties, both theoretically and experimentally,
to provide a clear and simple picture of the device behavior.
The experiment uses PMMA-passivated devices to reduce the hysteresis
seen in past work, allowing better comparison
with calculated characteristics.
The calculations use a simple quasi-classical
model to describe the diffusive transport of electrons and holes
along a long semiconducting nanotube.
This approach highlights the novel but simple physics
underlying the device operation.

\begin{figure}[ptb]
\includegraphics[height=3.5in,width=2.8in]{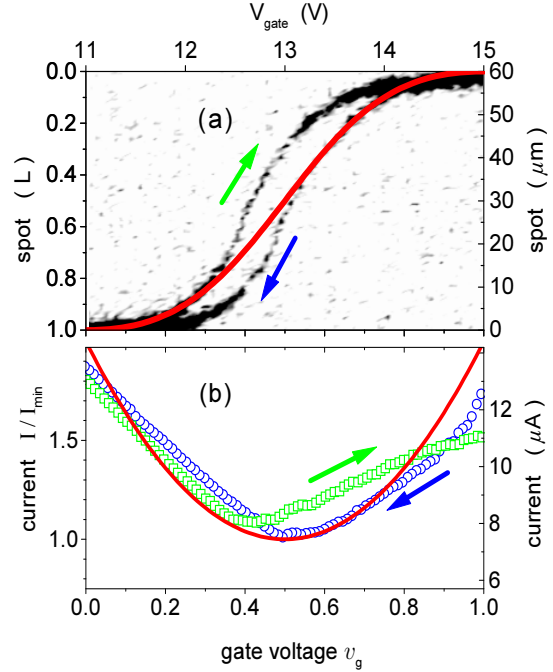}
\caption{Measured and calculated characteristics of electroluminescent CNFET.
(a) Position of light-emission spot (from drain electrode), and
(b) drain current $I$, vs gate voltage $V_{g}$.
Experiment is for $V_{d}=15V$.
Top and right scales show absolute units for experimental data.
Left and bottom scales show scaled units for comparison with theory.
Solid red curves are theory, Eqs.~(\ref{current2}-\ref{spot2}).
Experimental data is shown as raw image in (a); a video
is also available \cite{epaps}.  Data for current in (b)
are shown as open circle and squares.
Some hysteresis is visible between forward and reverse $V_{g}$ sweeps
in (a) and (b);
direction of each sweep is indicated by an arrow.
(There is also an unknown overall shift of the gate voltage scale
due to charge trapped in the oxide.)
\label{expmodel}
}
\end{figure}

We predict a ``universal'' behavior,
with \textit{no adjustable parameters} other than the overall
voltage and current scale. This prediction is directly compared
with experimental data for the motion of the light-emitting spot,
and for the electrical characteristics.
The agreement in Figure~\ref{expmodel} is striking,
in light of the simplicity of the model,
and the residual hysteresis in the experimental data.

This gives us confidence to apply the model, to extract information about the
transport, the field distribution, and the performance improvements available
by oxide scaling. We find that there is a strong enhancement of the electric
field near the emission spot, consistent with experimental observations
\cite{Fre04a,Fre04b}. The device performance is predicted to improve
dramatically with reduced gate-oxide thickness, due to both improved
gate-channel capacitance and reduced voltage drop at the contacts.

Single-wall carbon nanotubes are grown by chemical vapor deposition
\cite{Hua03, Fre04b} on degenerately doped silicon with 100nm silicon oxide.
CNFETs are fabricated as described in \cite{Fre04b}. The channel length is
$\sim$60 $\mu$m. We observe a strong hysteresis in the as-fabricated devices
\cite{Fre04b} that can be attributed to the trapping and de-trapping of
charges in the gate insulator and/or at states at the SiO$_{2}$ surface close
to the carbon nanotube. To reduce this effect,
we cover the device with PMMA and bake it out on a hot plate in air for 24h at
170C \cite{Kim03}.
This procedure
greatly reduces the hysteresis observed in the as-fabricated devices. There is
apparently still some charge in the oxide, but its effect can be approximated
as an overall shift of the $V_{g}$ scale. We measure the localized infrared
emission at $\lambda$=2.0 $\mu$m during gate-voltage sweeps with a liquid
nitrogen-cooled HgCdTe detector array, mounted on top of the camera port of an
optical microscope with 50x NIR objective lens \cite{Fre04b}.

We develop an explicit analytic description of the device properties,
based on the physical picture outlined in Ref.~\cite{Fre04b}.
Detailed simulations by Guo and Alam \cite{Guo05}
support this qualitative picture.
To model diffusive transport along semiconducting nanotubes, we take the
current to be%
\begin{equation}
I=-m\eta\left(  x\right)  \frac{dV\left(  x\right)  }{dx} \label{I}%
\end{equation}
where $x$ is the direction along the tube (from drain to source), $m$ is the
carrier mobility, $\eta$ is the linear density of carriers, and $V$ is the
electrostatic potential. Charge conservation in the steady state requires that
$I$ is the same for all $x$, so $\eta$ and $V$ cannot vary independently.

Because the gate oxide thickness is much less than the channel length, the
nanotube is very effectively screened by the gate. Therefore the potential on
the tube is determined by the gate voltage and the local charge \cite{Fre04b},
and can be approximated as
\begin{equation}
V\left(  x\right)  =V_{g}+C^{-1}\rho\left(  x\right)  \label{VC}%
\end{equation}
where $V_{g}$ is the gate voltage, $C$ is the nanotube capacitance per length,
and $\rho$ is the linear charge density.

Several aspects of the real system are simplified or neglected here. In
particular, we use the classical capacitance, neglecting the variation in
chemical potential with charge density.
And we treat the potential as a strictly local function of charge density,
which is only valid on length scales larger than the gate-oxide thickness. We
also neglect the dependence of mobility on electric field and charge density.
It is possible to avoid some of these simplifications in numerical
simulations \cite{Guo05}; and though performed in a relatively short-channel
regime, those simulations give confidence in the appropriateness of our
simplified treatment.

In keeping with our focus on long length scales, we also assume a negligible
recombination length, so at any $x$ there can be electrons or holes but not
both. Then $\rho=\pm\eta$ (positive for hole region), and combining
Eqs.~(\ref{I}) and (\ref{VC}) gives
\begin{equation}
\frac{dV\left(  x\right)  }{dx}=\mp\frac{I}{mC\left[  V\left(  x\right)
-V_{g}\right]  }\label{DiffEq}%
\end{equation}
This equation can be solved analytically in each region. The boundary
conditions on $V\left(  x\right)  $ are $V_{s1}$ and $V_{d1}$ (the potential
on the nanotube at the source and drain ends), and $V=V_{g}$ at the point
separating electron and hole regions. We focus on the ambipolar regime where
light emission is possible, which for $I>0$ is $V_{d1}\geq V_{g}\geq V_{s1}$.
Then
\begin{equation}
V\left(  x\right)  =V_{g}\pm\left(  \frac{2I}{mC}\right)  ^{1/2}\left\vert
x-x_{0}\right\vert ^{1/2}\label{Vx}%
\end{equation}
where $\pm$ is $-1$ for electrons and $+1$ for holes. Also
\begin{equation}
x_{0}-x_{d}=L\left[  \frac{\left(  V_{d1}-V_{g}\right)  ^{2}}{\left(
V_{s1}-V_{g}\right)  ^{2}+\left(  V_{d1}-V_{g}\right)  ^{2}}\right]
\label{spot}%
\end{equation}
where $x_{0}$ is the point separating electron and hole regions, and
$L=x_{s}-x_{d}$ is the channel length, $x_{s}$ and $x_{d}$ being the positions
of the source and drain contacts, respectively. Combining these gives%
\begin{equation}
I=\frac{mC}{2L}\left[  \left(  V_{s1}-V_{g}\right)  ^{2}+\left(  V_{d1}%
-V_{g}\right)  ^{2}\right]  \label{current}%
\end{equation}
in the ambipolar regime.

The equations above refer only to diffusive transport along the channel, and
do not include contact effects. Electrical contacts to nanotubes are a rich
subject in their own right --- the transmission depends on both the Schottky
barrier height and the electrode geometries \cite{Heinze02,Heinze03}, and no
doubt also on other factors that are less well understood. In general there is
a voltage drop $V_{c}$ associated with each contact, in addition to the
continuous voltage drop along the tube. For the ambipolar case there is
necessarily a voltage drop, equal to the bandgap, associated with the
crossover from electron to hole conduction, which in our model means that
$V_{c}$ cannot be less than the Schottky barrier height. Because of the highly
nonlinear behavior of Schottky contacts, we treat $V_{c}$ as constant here
(and assumed to be the same for both contacts) in the ambipolar regime,
analogous to a threshold voltage \cite{Rad03}.

Then following the usual convention that $V_{s}=0$ and $V_{d}$ denotes the
applied drain voltage, we have $V_{s1}=V_{c}$ and $V_{d1}=V_{d}-V_{c}$ in the
ambipolar regime. Combining this with Eqs.~(\ref{spot}-\ref{current}), we can
describe the behavior in a scale-free way, as%
\begin{align}
\frac{I}{I_{\text{min}}} &  =2\left[  v_{g}^{2}+\left(  1-v_{g}\right)
^{2}\right]  \label{current2}\\
\frac{x_{0}-x_{d}}{L} &  =\frac{\left(  1-v_{g}\right)  ^{2}}{v_{g}%
^{2}+\left(  1-v_{g}\right)  ^{2}}\label{spot2}%
\end{align}
Here $V_{d}-2V_{c}$ is the range of gate voltage over which the emission spot
sweeps from source to drain; $v_{g}=\left(  V_{g}-V_{c}\right)  /\left(
V_{d}-2V_{c}\right)  $ is the gate voltage as a fraction of this total sweep
voltage; and $I_{\text{min}}=\left(  mC/4L\right)  \left(  V_{d}%
-2V_{c}\right)  ^{2}$ is the minimum current, which occurs at $v_{g}=1/2$.

Our device measurements and model calculations
are summarized in Figure~\ref{expmodel}.
(A movie of the experimental spot motion
is also available \cite{epaps}.)
Figure~\ref{expmodel}b shows the gate
voltage characteristic of the device and
Figure~\ref{expmodel}a shows the corresponding movement
of the light-emission spot along the carbon nanotube. The experimental
measurements can be put directly on the same dimensionless scale as the model,
by noting that $v_{g}=0$ and $1$ are the gate voltages at which the spot
reaches the source and drain.
The overall agreement between the measured characteristics
and model calculations in Figure~\ref{expmodel} is striking. The spot
moves rapidly with $V_{g}$ when in the middle of the channel, and more slowly
near the source and drain. The shape of the spot-motion curve
is very well reproduced.
The current variation is also rather well reproduced,
to the extent of capturing
the symmetric minimum and general magnitude of variation.
The remaining discrepancies are attributable in part
to the residual hysteresis seen in the experiment.
However, the approximations of constant mobility
and constant voltage drop at the contacts also
play a role, we expect.

The experimental data also provide a direct determination of $V_{c}$. In
Fig.~\ref{expmodel}a, $V_{d}=15$V,
while the spot sweeps from source to drain over a
$V_{g}$ range of $\approx4$V. Equating this range with $V_{d}-2V_{c}$ suggests
that $V_{c} \sim 5-6$V. This is consistent with typical threshold voltages
observed for nanotube Schottky-barriers transistors on such thick gate oxides.

\begin{figure}[ptb]
\includegraphics[height=3.7in,width=2.9in]{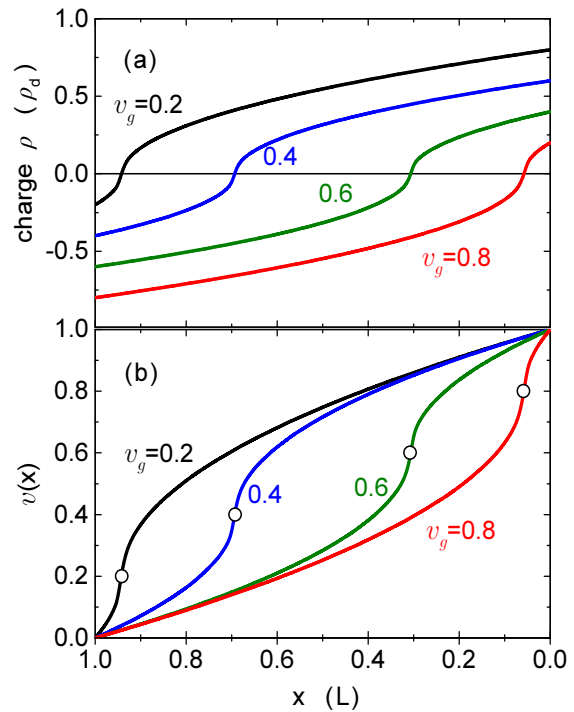}
\caption{ (a) Charge
density [Eq.~(\ref{rho})], and (b) electrostatic potential [Eq.~(\ref{Vx})],
along the nanotube (as fraction of channel length $L$, with drain at right),
for fixed drain voltage $V_{d}$ and different values of scaled gate voltage
$v_{g}$. Here $v(x)=(V(x)-V_{c})/(V_{d}-2V_{c})$. Curves from left to right
correspond to $v_{g}=0.2$, 0.4, 0.6, and 0.8. Circles indicate where $v=v_{g}$
and $\rho=0$. }
\label{vrho}
\end{figure}

Combining Eqs.~(\ref{VC}), (\ref{Vx}), and (\ref{current}), and incorporating
$V_{c}$, gives the charge density,
\begin{equation}
\rho\left(  x\right)  =\pm\rho_{d}\left[  \left(  v_{g}\right)  ^{2} + \left(
1-v_{g}\right)  ^{2}\right]  ^{1/2} \left\vert \frac{x-x_{0}}{L}\right\vert
^{1/2}\label{rho}%
\end{equation}
where $\rho_{d}=C\left(  V_{d}-2V_{c}\right)  $.
Figure~\ref{vrho} shows the calculated
charge and electrostatic potential along the tube for different values of the
gate voltage.
Within our model, the carrier density goes to zero at the
ambipolar emission spot, while the electric field driving the current diverges
there, as%
\begin{equation}
\frac{dV}{dx}=-\left(  \frac{I}{2mC}\right)  ^{1/2}\left\vert x-x_{0}%
\right\vert ^{-1/2}%
\end{equation}
Of course, in reality there is some overlap of electron and hole regions due
to the finite recombination rate; so this singularity is smoothed into a peak.
Still, the enhancement of electric field in the recombination region has
important consequences. It leads to strong hot-carrier effects in
electroluminescence \cite{Fre04a}, and increases the likelihood of Zener
tunneling at defects \cite{Fre04b}.

Our results suggest that such electroluminescence devices can be dramatically
improved by using thinner oxides, and further improved by using high-k
dielectrics. A major limiting factor is the high voltage required. But most of
the voltage drop apparently occurs at the contact. The threshold voltage for
the contacts scales with oxide thickness, and can be further improved by
optimizing the contact geometry and using different gate dielectrics
\cite{Heinze02,Heinze03}. Thus it should be possible to reduce the voltage
drop $V_{c}$ by an order of magnitude. The voltage drop along the channel
scales as $C^{-1}$, so this can also be greatly reduced with a thin high-k
gate dielectric. In this way the device could operate at lower voltages and
higher currents, give brighter and more efficient light emission.

We acknowledge Q. Fu and J. Liu for providing us with long CVD-grown carbon
nanotubes, J. Chen for helpful discussions, and B. Ek for expert technical assistance.

\end{document}